\documentclass[9pt,twocolumn]{article}

\usepackage{geometry}
\usepackage{graphicx}
\usepackage{multirow}
\usepackage{amsmath}
\usepackage{xcolor}
\usepackage{enumitem}
\usepackage{ragged2e}
\usepackage{fancyvrb}

\newcommand{\q}[1]{``#1''}

\usepackage{authblk}

\title{Data
Engineering for Data Analytics: \\ A Classification of the Issues, and
Case Studies}

\author[1]{Alfredo Nazabal}
\author[12]{Christopher K.I. Williams}
\author[3\footnote{The work described in this paper was carried out when Giovanni Colavizza and Angus Williams
were with the Alan Turing Institute}]{Giovanni Colavizza}
\author[1]{Camila Rangel Smith}
\author[*]{Angus Williams}
\affil[1]{The Alan Turing Institute, London, UK}
\affil[2]{School of Informatics, University of Edinburgh, UK}
\affil[3]{Media Studies Department, University of Amsterdam, The Netherlands}

\begin{document}

\maketitle

\begin{abstract}
Consider the situation where a data analyst wishes to carry out an
analysis on a given dataset. It is widely recognized that most of the
analyst's time will be taken up with \emph{data engineering} tasks
such as acquiring, understanding, cleaning and preparing the data. In
this paper we provide a description and classification of such tasks
into high-levels groups, namely data organization, data quality and
feature engineering.  We also make available four datasets and example
analyses that exhibit a wide variety of these problems, to help encourage
the development of tools and techniques to help reduce this burden and push forward research towards the automation or semi-automation of the data engineering process.
\end{abstract}


%

\section{Introduction}\label{sec:introduction}

%
%
%
%

A large portion of the life of a data scientist is spent acquiring,
understanding, interpreting, and preparing data for analysis, which we
collectively term \emph{data engineering}\footnote{Another term
often used is data wrangling.}. This can be time-consuming and
laborious, for example \cite{dasu-johnson-03} estimate that these
tasks constitute up to 80\% of the effort in a data mining project.
Every data scientist faces data engineering challenges
when carrying out an analysis on a new dataset, but this work is often
not fully detailed in the final write-up of their analysis.

Our focus is on a task-driven analysis scenario, where a data
scientist wants to perform an analysis on a given dataset, and they
want to obtain a representation of the data in a format that can be
used as input to their preferred machine learning (ML) method.
Interestingly, there is no unique or correct way of wrangling a
messy dataset.
For example, features and
observations with missing values can be deleted or imputed using
different approaches; observations can be considered anomalous
according to different criteria; or features can be transformed in
different ways.

Furthermore, most publicly available datasets have already undergone
some form of pre-processing. While providing a clean version of the
data is
extremely helpful from a modelling perspective, data engineering
researchers suffer from a limited availability to public messy
datasets. This leads to researchers addressing some of the problems by
synthetically corrupting clean datasets according to the data
wrangling problem to be solved. However, such
synthetic corruption is often not sufficient to capture the 
wide variety of corruption
processes existing in real-world datasets. All of this variability in
terms of data engineering issues and possible solutions, and the lack of
public messy datasets and their final cleaned version makes attempting to automate this problem extremely challenging.

In this paper we first provide a classification of data engineering
problems appearing in messy datasets when a data scientist faces an
analytical task, and give practical examples of each of them.  We have
identified three high-level groups of problems: {\bf Data
  Organization} issues (DO), related to obtaining the best data
representation for the task to be solved, {\bf Data Quality} issues
(DQ), related to cleaning corrupted entries in the data, and {\bf
  Feature Engineering} (FE) issues, related to the creation of
derived features for the analytical task at hand.  Additionally, we have
further divided the DO and DQ groups according to the nature of data
wrangling problem they face. Under Data Organization we include data
parsing (DP), data dictionary (DD), data integration (DI) and data
transformation (DT). Under Data Quality we include canonicalization
(CA), missing data (MD), anomalies (AN) and non-stationarity (NS).
Providing a classification of wrangling challenges not only pushes
research in these individual fields, but also helps to advance the
automation or semi-automation of the whole data engineering process.

A second contribution of the paper is to make available four example
messy datasets, each with an associated analytical task. The analyses
were carried out by data scientists at the Alan Turing Institute (in
some cases replicating steps that were taken for published
analyses). In the appendices we describe the cleaning operations they
needed to perform in order to obtain a version of the data that could
be used for the analysis task at hand. These provide practical
examples of the issues identified in our classification, and give an
insight of what constitutes a practical pipeline during a data
wrangling problem.


The structure of the paper is as follows: Section \ref{sec:case-studies}
introduces the four cases studies that will provide examples
throughout the paper. Section \ref{sec:class} describes our
classification of data engineering challenges, broken down under the
headings of Data Organization, Data Quality and Feature Engineering.
Section \ref{sec:rel-work} discusses related work. Section \ref{sec:examples} provide details of the data engineering and modelling steps carried out for each of the four datasets, and our
conclusions appear in section \ref{sec:conc}.

\section{Overview of Case Studies \label{sec:case-studies}}

We have identified four case studies, drawing data from a variety of
domains and having a variety of formats: plant measurements, household
electricity consumption, health records, and government survey
data. We refer to Section~\ref{sec:examples} for detailed descriptions of the data engineering challenges and modelling approaches of each dataset. We also provide a
GitHub repository with the datasets and the wrangling processes.\footnote{https://github.com/alan-turing-institute/aida-data-engineering-issues}
%
%
%
Table~\ref{t:AIDA_problems} shows an overview of the wrangling challenges present in each of
our use cases. Note that the actual challenges that an analyst
encounters depend on the particular analysis that is undertaken, which
in turn will depend on the particular analytical task being addressed.
\justify
{\bf Tundra Traits dataset}: The Tundra Traits data consists of measurements of the physical characteristics of shrubs in the Arctic tundra, as well as the data wrangling scripts that were used to produce a \q{clean} version of the dataset~\cite{bjorkman-myers-smith-etal-18}. The analytical challenge entails building a model to gauge the effect of temperature and precipitation on specific shrub traits related to plant growth (see Section~\ref{sec:Tundra}).
\justify
{\bf Household Electricity Survey (HES) dataset}: The Household Electricity Survey (HES) data contains time-series measurements of the electricity use of domestic appliances, as well as a report on data cleaning by Cambridge Architectural Research~\cite{HES}. Our chosen analytical challenge is to predict the energy consumption of a household given its appliance profile (see Section~\ref{sec:HES}).
\justify
{\bf Critical Care Health Informatics Collaborative (CleanEHR)
  dataset}: The CleanEHR data contains a sample of anonymized medical
records from a number of London hospitals, including demographic
data, drug dosage data, and physiological time-series
measurements~\cite{cleanEHR}, together with publicly available data
cleaning scripts.
Our chosen analytical challenge is to predict which patients die in the first 100 hours from admission to the hospital. See Section~\ref{sec:cleanEHR}.
\justify
{\bf Ofcom Consumer Broadband Performance dataset}: The Consumer
Broadband Performance dataset contains annual surveys of consumer
broadband speeds and other data, commissioned and published by
Ofcom\footnote{The UK Office of Communications.}, and available as a
spreadsheet for each year~\cite{Broadband}. The data engineering
challenge here is primarily one of matching common features between the
data for any two years, as the names, positions, and encodings
change every year.
There was no given analytical challenge with the data as published, so we have chosen to build a model to predict the characteristics of a region according to the properties of the broadband connections (see Section~\ref{sec:broadband}).
\begin{table}[h]
\setlength{\tabcolsep}{2pt}
\centering
\caption{List of wrangling issues encountered in the datasets.}
\label{t:AIDA_problems}
\vspace{0.3cm}
\begin{tabular}{|l|cccc|cccc|c|}
\cline{2-10}
\multicolumn{1}{l|}{} & \multicolumn{4}{|c|}{Data} & \multicolumn{4}{|c|}{Data} & Feature \\
\multicolumn{1}{l|}{} & \multicolumn{4}{|c|}{Organization} & \multicolumn{4}{|c|}{Quality} & Engineering \\
\hline
\multicolumn{1}{|l|}{Dataset}            & DP & DD & DI & DT & CA & MD & AN & NS & FE \\
\hline
\multicolumn{1}{|l|}{Tundra}  &  & $\bullet$ & $\bullet$ & $\bullet$ & $\bullet$  & $\bullet$ & $\bullet$ & & $\bullet$ \\
\multicolumn{1}{|l|}{HES} & & $\bullet$ & $\bullet$ & $\bullet$ &  $\bullet$ & $\bullet$ & & & $\bullet$  \\
\multicolumn{1}{|l|}{CleanEHR} &  & $\bullet$ & $\bullet$ & $\bullet$ & $\bullet$ & $\bullet$ &  & $\bullet$ & $\bullet$\\
\multicolumn{1}{|l|}{Broadband} & $\bullet$ & $\bullet$ & $\bullet$ &  &  & $\bullet$ & & $\bullet$ & $\bullet$ \\
\hline
\end{tabular}
\end{table}

\section{Classification of Wrangling Challenges \label{sec:class}}

When data analysts want to employ their preferred machine learning
models, they generally need the data to be formatted in a single
table.  Although data engineering issues extend to other forms of data
such as images, text, time series or graph data, in this work we take
the view that the data analyst wishes to obtain a clean regular table
that contains the information needed to perform the desired analytical
task. To define notation, a table is assumed to consist of $n$ rows
and $D$ columns. Each row is an \emph{example} or \emph{instance},
while each column is a \emph{feature} or \emph{attribute}, and
attributes can take on a set of different \emph{values}.

We mainly focus on two large groups of data wrangling issues: those related with \emph{organizing} the data and those related with improving the \emph{quality} of the data. We also include a third group on \emph{feature engineering}, where the separation between wrangling issues and modelling choices starts to become blurred. In Table~\ref{t:summary} at the end of this section we provide a summary of all the data wrangling problems addressed in this work.
Notice that this classification should not be considered as a sequential process. Depending on the dataset and the analytical task some challenges are necessary while others are not, and the order of solving these problems  can be variable.

\subsection{Data Organization}

When data scientists start to work on a specific problem with real
data, the first problem they face is obtaining a representation or
view of their data that is best suited for the task to be solved. Data
scientists usually perform several operations while organizing their
data: the structure of the raw data is identified 
so that it can be read properly (data parsing); a basic
exploration of the data is performed, where basic metadata is inferred
for all the elements in the data (data dictionary); data from multiple
sources is grouped into a single extended table (data integration);
and data is transformed from the raw original format into the desired
data structure (data transformations).

\subsubsection{Data Parsing}

{Data parsing refers to the process of identifying 
the structure of the raw data source so that it can be read properly.}
Raw data sources come in many different formats, from
CSV files to flat files, XML, relational databases, etc. Some file
formats can be read using open source programs while others are only
readable using proprietary software.

Even when data comes in a well-structured file format, it can present
a wide variety of encodings. For example, the well-known CSV format
allows a wide variety of delimiter, quote and escape characters, and
these have to be determined before the file can be loaded into an
analysis program. Additionally, multiple tables may appear together in
a single file, making it necessary to identify them and extract them
properly.  Even though many CSV files can be easily read by standard
libraries, it is not uncommon to find files where this process fails,
and the analyst needs to manually select the parameters to properly
parse the data; see~\cite{vandenburg-nazabal-sutton-19} for
state-of-the-art performance on this task.
%
\begin{table}[h]
\centering
\caption{An example of an ambiguous CSV file from~\cite{vandenburg-nazabal-sutton-19}.}
\vspace{0.3cm}
\label{t:CSV_file}
\begin{tabular}{|l|}
\hline
\verb|Mango; 365,14; 1692,64| \\
\verb|Apple; 2568,62; 1183,78| \\
\verb|Lemon; 51,65; 685,67| \\
\verb|Orange; 1760,75; 128,14| \\
\verb|Maple; 880,86; 323,43| \\
\hline
\end{tabular}
\end{table}
\textit{Practical Example}: Table~\ref{t:CSV_file} shows an example of
ambiguous CSV files appearing in real-world
scenarios. A wrong
detection of the CSV parameters can lead to a completely different
table output (e.g.\ notice how choosing either semicolon, space or comma as the field separator results in different structured
tables with three columns).
%
%


\subsubsection{Data Dictionary}

The term \emph{Data Dictionary} refers to \emph{understanding} the
contents of the data and translating that knowledge into additional
metadata. Ideally a dataset should be described by a data dictionary
or metadata repository which gives information such as the meaning and
type of each attribute in a table, see
e.g.~\cite{abedjan2015profiling}. This information is normally
provided by the data collectors and may come in many different
formats, e.g.\ text documents with a profile of the data, extra
headers in the data with information about its features in the
data, or additional CSV files with descriptions for all the
features. When data scientists are working with the data is it common
that they need to go back and forth from the data to the metadata to
understand the meaning of the tables and features and to
detect possible inconsistencies that are not obvious from a simple
exploration of the data. Even more importantly, it is not uncommon
that this metadata is missing or out-of-date, making it necessary to
infer it from the data itself. This process of data dictionary
creation takes place at the level of tables, features and feature
values:
\justify
{\bf Table understanding}: The first problem data scientists face is
to understand what is the global information contained in the tables,
or \emph{what is this table about?}. Often the owners of the data
provide this information directly to the data scientists, or in some
cases the name of the files reflect broadly the information contained
in the data, but this is not always so.  Data scientists then proceed
to \emph{explore} the data and try to understand their contents, by
answering questions such as: How many tables are contained in the
data? How many features? How many instances? Is there metadata with
information about the features in the data?  Additionally,
understanding whether the data is a relational table, a time series or
another format is a key process to decide how to work with this data.
Unfortunately, data scientists might not be able to fully understand
the contents of the data simply by exploration (e.g.\ headers might be
missing, file names might be non-informative, etc.). In such cases,
table understanding involves an interaction with a domain expert or
the data collectors.

\textit{Practical example}: In the Broadband dataset all basic information about the tables is summarized in the names of the different directories that contained the
dataset. Each folder name contained the year and month of each annual
survey. Most of the time, each folder only contained a CSV file with the different measurements for that year (panellist data). However, some of these folders also contained an additional CSV file (chart data) with a collection of tables and captions for those tables, with some basic summaries about the complete dataset.
\justify
{\bf Feature understanding}: Normally, some basic information on
the features contained in the table is provided, either by a header
giving the name of each feature, or by an additional file with detailed
information of each feature. When this information is not provided,
either a domain expert should be involved to help complete the missing
information, or this information must be inferred directly from the
data. Feature understanding can take place at a semantic level
(e.g.\ a feature with strings representing countries, see
e.g.\ \cite{chen-jimenez-ruiz-horrocks-sutton-19}), or at a syntactic
level (e.g.\ a feature is an integer, or a date~\cite{ceritli2019ptype}). Feature understanding is a crucial process that must be addressed when working with a new dataset. Depending on the predictive task, the data scientist might need to remove features that leak information on the target feature, remove unnecessary features for the task at hand, transform some of these features, or create new features based on existing ones. This process relates heavily with \emph{Feature Engineering} (see Section~\ref{sec:feat_eng}).


\textit{Practical example 1}: From the latitude and
longitude features in the Tundra dataset, we can get a sense of
the locations around the world where plant traits were measured. From
the experimental protocol (metadata), for each observational site there were a number of individual plots (subsites), and all plants within each plot were recorded. Unfortunately, there was no indication in the data whether
the latitude and longitude for each plant was recorded at the site or 
plot level. Further exploration of the data was needed to discover that the coordinates apply at site level.

\textit{Practical example 2}: In the CleanEHR demographics
dataset, the names of the features in the header were not informative
for a non domain expert, with names like DOB, WKG, HDIS or DOD. A
systematic matching process of each feature name in the dataset with a website containing the
descriptions and information of the different features was necessary to understand the data.
\justify
{\bf Value understanding}: Exploring the values contained in the different features give data scientists a better sense of what is contained in the data, and it is the first step towards fixing possible errors in the data. Some common questions data scientists normally try to answer during this process are: if a feature is categorical, how many unique elements does it have? Do some features contain anomalous values? Do values outside the expected range of values of a given feature exist? Value understanding is especially important during the feature understanding process when insufficient information about the features was provided. 

\textit{Practical example}: the \q{HouseholdOccupancy} feature in the HES dataset, which was inferred to mean \q{number of people living in the house} from its name, was not numeric. A value \q{$6+$} was used to encode that a household had six or more tenants. Since the data scientist wanted this feature to be numerical, these \q{$6+$} entries were transformed into \q{$6$}.


\subsubsection{Data Integration \label{sec:dataint}}
\label{sec:integration}

Data integration involves any process that \emph{combines}
conceptually related information from multiple sources. Data
analysts usually try to aggregate all the information they need for
their task into a single data structure (e.g.\ a table, or time
series, etc.), so they can use their desired machine learning model for the predictive task. However, data is often received in installments,
(e.g.\ monthly or annual updates), or separated into different tables
with different information and links between them
(e.g.\ demographic information about patients, and medical tests
performed on each patient.). Additionally, it is not uncommon that
the structure of the data may change between installments (e.g.\ an
attribute is added if new information is available). Commonly, data
integration consists of either \emph{joining} several tables together, i.e.,
adding new features to an existing table (usually referred as record linkage), or \emph{unioning} several tables, i.e., adding new rows to an existing table. Further
issues arise when the original data sources contain different data
structures (e.g.\ relational tables and time series).
\justify
{\bf Record linkage and table joining}: This process involves
identifying records across multiple tables that correspond to the same
entity, and integrating the information into a single extended table.
(Record linkage is also called entity disambiguation,
or record matching \emph{inter alia} \cite{elmagarmid2006duplicate},\cite{christen2011survey}.)  
If \emph{primary keys} exist in the data, the process is relatively
straightforward.  In such cases, joining two or more tables is usually
performed by identifying the common keys across tables and creating an
extended table with the information of the multiple tables. However, more
involved linkage processes are common during data integration: the name of
primary keys can change across tables, a subset of features in
the data might be needed to perform the record linkage (e.g.\ one
feature identifies a specific table and another the entity (row) in
that table), or even more specialized mappings between features might
be needed (e.g.\ in Tundra Traits as per Practical example 2 below).
%
The record linkage process is even harder when explicit
keys are not available in the data. In such cases, a
\emph{probabilistic} record linkage problem arises. A broad subset of
potential features are assigned with some weights that are used to
compute the probability that two entities are the same. The
probability of a pair of entities exceeding a given threshold is used
to determine whether they are a match or not. Early work on a 
theoretical framework for the problem was developed by Fellegi and
Sunter~\cite{fellegi1969theory}. See \cite{sayers2015probabilistic} and \cite{murray2015probabilistic} for more recent work on probabilistic record linkage.

\textit{Practical example 1}: The HES dataset is contained in 36 CSV files spread across 11 directories. Although several keys appear repeatedly in different tables, unfortunately, the names of these keys were not always the same across tables, even when they refereed to the same concepts. For example, the household identifier was sometimes given the name \q{Household} while in other tables was called \q{Clt Ref}. Additionally, each key contained different information regarding the combination of different sources. For example, \q{Household} identified the specific household for which monitoring took place (ID feature), \q{appliance code} was a number
corresponding to a specific type of appliance (e.g.\ 182 $\rightarrow$
trouser press) and \q{interval id} was a number referring to the type of
monitoring for a given appliance (e.g.\ 1 for monitoring conducted
over one month at two minute intervals).

\textit{Practical example 2}: The Tundra dataset presents an
interesting table joining problem. It was necessary to combine the \q{Tundra Traits Team database}, which
contained different measurements from plants across the world, with
the \q{CRU Time Series temperature data and rainfall data} datasets,
which contained the temperature and rainfall data across the world,
gridded to 0.5º, with a separate table for each month between 1950 and 2016. This process was needed since
temperature and rainfall have an effect on the measured traits of the
plants. The joining of both data sources was performed with a two step process for each plant:
\begin{enumerate}
\item Identify the year and month when the specific trait for a plant was recorded, and look for the corresponding table into the CRU catalog.
\item Check the longitude and latitude values for that plant in the Tundra dataset, and map the corresponding temperature from the CRU dataset into the extended table.
\end{enumerate}
Four different keys were thus necessary in the Tundra dataset (\q{DayOfYear}, \q{Year}, \q{Longitude} and \q{Latitude}) to connect the temperature data in the CRU catalog into the extended table.
\justify
{\bf Table unioning}: This process involves aggregating together row-wise different tables that contain different entities with the same information into one extended table. While ideally concatenating several tables together should be straightforward, in many real world scenarios the structure of different tables can differ dramatically. Some common problems that arise during table unioning include: features located
in different positions in two or more tables, header names changing across
tables, or features being added or deleted. To address all these problems, the authors in \cite{sutton-hobson-geddes-caruana-18} obtained a matching between the attributes in two datasets, based on the idea that the statistical \emph{distribution} of an attribute should remain similar between the two datasets.

\textit{Practical example 1}: In the Broadband dataset, the different annual
installments distributed across different CSV files needed to be concatenated into a single
extended table. An exploration of each table reveals several problems: some columns were renamed between
tables, the order of the columns was not preserved, not all columns
appeared in each table and some categorical columns were recoded.
The union process for all these tables involved selecting a common
structure for the extended dataset, and mapping every feature from each
table into the corresponding feature of the extended table. Doing this
process manually (instead of using the program in \cite{sutton-hobson-geddes-caruana-18})
involved accounting for every single particular mapping across tables in the code.
\justify
{\bf Heterogeneous integration}: This process involves combining data
from different structured sources (relational tables, time series,
etc) and different physical locations (e.g. websites, repositories in different directories, etc.) into a single extended structure format. Data
repositories can be massive, consisting of multiple sources across
different locations. Getting all this information into	 a proper
format is time consuming and sometimes not feasible. Since the data
scientists might be interested only in extracting a small subset of
the data that refers to their particular problem, query based systems
are needed during the integration process. This parsing is often
handled by an ETL (Extract, Transform and Load)
procedure~\cite{vassiliadis2009survey}, but such code may be
difficult to write and maintain. The idea of Ontology Based Data
Access (OBDA; \cite{poggi-etal-08}) is to specify at a higher level
what needs to be done, not exactly how to do it, and to make use of
powerful reasoning systems that combine information from an ontology
and the data to produce the desired result.

\textit{Practical example}: The CleanEHR dataset is composed of a
collection of R objects that contained demographic information about
the different patients, time series measurements of the different
medical tests that were performed on the patients and some CSV files
with detailed information about different encodings employed in the
data. Since the modelling task is related to the times of death of
the patients, only parts of the time series data are needed to train
the predictive models, since in a real predictive scenario future
information is not available. With these considerations, only the
files containing the first 10 hours of the patients' stay in the
hospital would be extracted to train the model, discarding all the patients that lived less
than 10 hours, in order to include as much data as possible. From
those selected files, only relevant summary measures from the
time-series data (mean, standard deviation and last value measured)
should be combined with the demographics data during the table joining process.
\justify
{\bf Deduplication}: It is common after a data integration process
that several duplicate instances containing the same or similar information exist
in the extended table~\cite{christen2011survey}; these duplicates need to be resolved into one instance. Note that, while record linkage involves identifying matching
entities across tables to create an extended table with all the
available information, deduplication involves identifying matching
entities in a specific table to remove the duplicate entries in the
data. Similar techniques employed during the matching process in
record linkage can also be employed during deduplication\footnote{The 
generic term \emph{entity resolution} encompasses both 
record linkage and deduplication.}.


\subsubsection{Data Transformation}
\label{sec:data_transformation}

Most of the time, data analysts need the data to be formatted in a
$n \times D$ table, where every instance should have $D$ attributes.
However, the original shape of the data does not always conform to
this structure,
and one may need
to transform the \q{shape} of the data to achieve the desired form~\cite{wickham2014tidy}. Furthermore, the original data source may not be tabular, but unstructured or semi-structured. In this case we need to carry out \emph{information extraction} to put the data into tabular form.
\justify
{\bf Table transformations}: This process involves any manipulation on the data that changes
the overall \q{shape} of the data. One of the most common table
manipulations is the removal of rows or columns. There
are many reasons why data scientists remove parts of the data
(e.g.\ some rows or features being completely or almost completely
empty, some information is deemed unnecessary for the analysis task,
etc.). Another common table manipulation is switching the format of
the table from a \q{wide} to a \q{tall} format or vice versa~\cite{wickham2014tidy}, which can happen e.g.\ when column headers in the data are values and not variable names. Note that the \q{tall} format is closely related to the one encoded in a semantic (or RDF) triple, which has the
form of a (subject, predicate, object) expression; for example
(patient 42, has height, 175cm).

\textit{Practical example 1}: In the extended table of the Tundra dataset scenario, from all the possible plant traits, only a small subset
of them were measured for each plant. In the original wide format (see Section~\ref{sec:integration}, record linkage practical example 2), the table is extremely sparse, with many missing entries. By changing the table from a \q{wide} to a \q{tall} format and grouping all the trait names under a
new feature \q{Traits} and the values under a new feature \q{Values}, a more compact table is created with the missing trait values
not being explicitly represented. 
Additionally, the data scientist removed all observations where: a) a
geographical location or year of observation was not recorded, b) a
species had less than four observation sites and c) sites had less
than 10 observations. Notice the new structure of the table was
decided according to the analytical task
at hand (other structures are possible).
\justify
{\bf Information extraction}: Sometimes the raw data may be
  available in unstructured or semi-structured form, such as free
  text.  A process involving the extraction of the relevant pieces of
  information into multiple features, while discarding any unnecessary
  information is often required. Examples include named entity
  recognition~\cite{nadeau2007survey} (e.g.\ names of people or places), or relationship 
extraction~\cite{leng2016deep} (e.g.\ a person is in a particular location). Natural 
language processing (NLP) methods are often used to tackle such
tasks.

\textit{Practical example}: 
FlashExtract by Le and Gulwani \cite{le2014flashextract} provides a 
programming-by-example approach to information extraction. A file
contains a sequence of sample readings,
where each sample reading lists various \q{analytes} and their
characteristics. By highlighting the different required target fields 
for a few samples, a program is synthesized (from an appropriate
domain specific language) which can extract the required information
into tabular form.


%

\subsection{Data Quality}

When the data has been properly structured into the desired output
format, data scientists need to carefully check it in order to fix any
possible problems appearing in the data. Data quality issues involve any
process where data values need to be modified without changing the
underlying structure of the data. Common data cleaning operations
include standardization of the data (canonicalization), resolving
missing entries (missing data), correcting errors or strange values
(anomalies) and detecting changes in the properties of the data
(non-stationarity).


\subsubsection{Canonicalization \label{sec: canon}}

Canonicalization (aka standardization or normalization) involves any
process that \emph{converts} entities that have more than one possible
representation into a \q{canonical} format. We can identify different canonicalization problems in datasets according to the transformations needed to standardize the data. One form of canonicalization involves obtaining
a common representation for each unique entity under a specific
feature (e.g.\ U.K., UK and United Kingdom refer to the same entity
under different representations.). This process usually involves a
previous step of identifying which values of a given feature
correspond to the same entities, which  we term \emph{cell entity resolution}. Another form of
canonicalization involves obtaining a standard representation for
every value under a given feature (e.g.\ specific formats for dates,
addresses, etc.). This standard representation is often chosen by the data scientist according to the task they intend to solve. A particular case of a feature canonicalization involves standardizing the units of physical measurements across a given feature. The final goal of a canonicalization process is to obtain a table where every entity in the data is represented uniquely and all the features in the data follow a common format.
\justify
{\bf Cell entity resolution}: This process involves identifying when
two or more instances of a given feature refer to the same
entity. This process has many similarities with deduplication and
record linkage (see Section~\ref{sec:dataint}). The main difference
is that the goal of deduplication and record linkage is to match
complete instances across the same table or across multiple tables,
while cell entity resolution is only concerned about identifying when
two or more instances with a different representation in a given
feature refer to the same entity. A lot of variability can arise when
inputting values into a dataset. Different names for the same entity,
abbreviations, or syntactic mismatches during recording (e.g.\ typos,
double spaces, capitalization, etc.) lead to many different entries in
a feature referring to the same underlying concept. Identifying which
cells correspond to the same entity allows to repair them by matching
them to a common or standard concept\footnote{
We note that 
\cite{cerda-varoquaux-kegl-18} argue that such cell entity resolution
may not be necessary if one uses a similarity-based encoder for 
strings which gracefully handles  morphological variants like typos.
This may indeed handle  variants like U.K.\ and UK, but it seems
unlikely to group these with ``United Kingdom'' unless some knowledge 
is introduced.}. The software tool OpenRefine
\cite{openrefine20} provides some useful functionality (``clustering'')
for this task.

\textit{Practical example}: In the Tundra dataset there are two features referring to the names of the plants in the dataset: \q{AccSpeciesName} and \q{OriginalName}. While \q{AccSpeciesName} contains the name of the plant species, the \q{OriginalName} contains different names that the scientists employed to record those plants, including typos, capitalization changes, etc.  (e.g.\ AccSpeciesName: \q{Betula nana} - OriginalName: [\q{Betula nana}, \q{Betula nana exilis}, \q{Betula nana subsp. exilis}]).
\justify
{\bf Canonicalization of features}: This refers to any process that involves
\emph{representing} a specific feature type (e.g. dates, addresses, etc) with a standard
format. For example, phone numbers can be encoded as \q{(425)-706-7709}
or \q{416 123 4567}, dates can be represented as 25 Mar 2016 or
2016-03-25, etc. While different feature formats exist for a
specific type, is it a decision of the data scientist to 
select the desired standard format for a feature (usually related
to the task to be solved) and convert all the values in that feature
to that standard format. This can often be done by writing regular
expressions, but this is only possible for skilled
operators. The authors in \cite{gulwani-11} developed a method for \emph{learning}
these transformers via programming-by-example. This was later made
available as the Flash Fill feature in Microsoft Excel to a very wide
user base.

\textit{Practical example:} Features related with dates and time in
the CleanEHR dataset occur in different formats. 
For example in one case both the date and time were
recorded in the same feature (DAICU), and in another case date and time
were recorded under two different features (DOD and TOD). Each feature
needed to be dealt case-by-case by the data scientist and transformed
into a common datetime Python object.
\justify
{\bf Canonicalization of units}: This refers to any process that involves \emph{transforming} the numerical values and units of a feature into a standard representation. This problem commonly arises from physical measurements being recorded with different units. For
example, height can be recorded both in metres or centimetres,
temperature in Celsius or Fahrenheit, etc. Different
unit formats can lead to different representations appearing in the data, and the data scientist needs to standardize all of them into a single representation.

\textit{Practical example 1}: A common unit canonicalization problem appears in the HES dataset. A feature in the dataset measured the volume of the refrigerators in each household (\q{Refrigerator volume}). Unfortunately, the values were recorded with different units (e.g.\ litres, cubic feet or no unit), which were also written in different formats (e.g.\ \q{Litres}, \q{Ltr},\q{L}). A process of removing the units from the feature and standardizing all the values according to a common unit (litres in this case) would be necessary.

\textit{Practical example 2}: Several of the features in the CleanEHR
dataset were recorded with different units. For example, the feature
\q{HCM} containing the height of the patients was supposedly recorded
in centimetres, but around 200 patients had their height
recorded in metres. Additionally, another feature
\q{apacheProbability}, recording the probability output of the apache test,
was recorded in the range $[0,100]$ in some cases and in other cases
between the range $[0,1]$. Notice that, in this case, the units were not
provided in the data, and reasoning about the meaning of the feature
and its values was necessary to determine that a unit canonicalization
problem was present for those features. According to the protocol
description, \q{HCM} was encoded in centimetres and
\q{apache-probability} in the $[0,1]$ range.


\subsubsection{Missing data}
\label{sec:Missing}

This issue concerns every operation that involves detecting,
understanding and imputing missing values in a dataset. Missing
data is a challenging problem arising very frequently in the data
wrangling process~\cite{schafer2002missing}. Since many machine
learning models assume that the data is complete, data scientists need
to detect the missing values and repair them properly. Otherwise, they would be restricted to employing prediction methods that can handle incomplete data.
\justify
{\bf Detection}: Datasets often contain missing
entries. These may be denoted as \q{NULL}, \q{NaN} or \q{NA}, but
other existing codes are widely employed like \q{?}, or
\q{-99}. While codes like \q{NULL} or \q{NaN} are easily
identified by a simple exploration of the data, other values
encoded with the same type of the feature (e.g.\ \q{-99} for numerical
values) are \emph{disguised missing values}~\cite{pearson2006problem}
and can only be identified by understanding the properties of the features.

\textit{Practical example}: At first glance, all the missing values in the CleanEHR dataset seemed to be encoded as \q{NaN}. However,
after further inspection of features \q{apache-probability} and
\q{apache-score}, the analyst noticed that two supposedly positive features
had a significant number of \q{-1} values. Since \q{-1} does not
conform with the range of possible values of those features, these codes were inferred to represent when the apache test was not provided, and consequently, they should be treated as missing values.
\justify
 {\bf Understanding}: The underlying process of the missing data patterns can provide additional information about the contents of the data. The book by 
\cite{little-rubin-87} is a seminal reference on the understanding and \emph{imputation} of missing data, classifying the missing patterns in three groups according to the generation process. a) Missing Completely At Random (MCAR), when the  pattern of missingness does not depend on the data, b) Missing At Random (MAR), when the missingness only depends on observed values and c) Missing Not At Random (MNAR), when the missingness can depend both on observed and missing values.

\textit{Practical example}: Different missing data patterns exist in the CleanEHR dataset, depending on the features. For example, the \q{HCM} feature seemingly had entries
missing with no correlation with other values in the data (MCAR),
while some of the medical tests (e.g.\ \q{CHEMIOX} or \q{RADIOX}) had
missing entries that seemed to indicate that a specific test
was not performed (MNAR). Unfortunately, these features also had
entries with \q{0} values, indicating clearly that a test was not
performed (see practical example 1 in Section~\ref{sec:non_stat}). From a data scientist's perspective, this means that we cannot  be sure whether a missing value indicates a test not performed, or whether
there were other reasons for the entry being missing (e.g.\ someone
forgot to fill this entry).
\justify
{\bf Repair}: This process involves any operation performed on the
missing data. The most common procedures are either removing any
rows with missing entries (known as \q{complete case}
analysis) or substituting those missing entries with other values
according to different rules (imputation). Some imputation methods include replacing
every missing entry in a feature by the mean or mode of the statistical
distribution (mean/mode imputation), filling those missing entries
with values sampled randomly from the observed values of the feature,
or applying more sophisticated models for missing data imputation
(see~\cite{schafer2002missing} for some classical methods
and~\cite{MICE11},\cite{yoon2018gain} or \cite{nazabal2018handling}
for recent work). Such multiple imputation
methods can represent the inherent uncertainty in imputing missing data.
We would like to remind the reader of the
difficulty of this problem in a real world scenario: ground truth for
the missing entries is not available, so a data scientist needs to
be sure that whatever method they decide to use during repair will
impute sensible values for those entries.

\textit{Practical example 1}: in the Tundra extended
table,  the geographical location, the date of the observation
and temperature features at a site were often missing. During the integration process, whenever either the location or the year was not provided, the
temperature could not be joined into the extended table, resulting in a missing
entry. Since these entries would become problematic for the modelling
task, every plant where any of these features were missing was removed from the dataset.

\textit{Practical example 2}: for the missing values in the HES dataset, different approaches were followed depending on the feature. For example, for the \q{House.age} feature, a unique missing entry with a code of \q{-1} exists, and it was imputed using the most common age range in the feature (mode imputation).

\textit{Practical example 3}: a similar procedure was also employed in the CleanEHR dataset. For example, the missing entries in \q{HCM}, most likely presenting a MCAR pattern, were imputed using the mean of the distribution (mean imputation) after standardizing the feature (converting the metres
values into centimetres, see canonicalization in Sec.\  \ref{sec: canon}).
 However, for the medical test (\q{CHEMIOX},\q{RADIOX}, etc.) those missing values were imputed with a \q{0},
according to the belief that a missing entry indicated that a test
was not performed.



\subsubsection{Anomalies}

An anomaly can be defined as \q{a pattern that does not conform
to expected normal behaviour} \cite{chandola2009anomaly}. Since anomalies
can potentially bias models to wrong conclusions from the data,
identifying and possibly repairing these values becomes critical
during the wrangling process. Anomalies arise in the data
for a variety of reasons, from systematic errors in measurement
devices to malicious activity or fraud. Anomaly detection is such an
important topic that in many problems the detection of anomalies in a
dataset is the analytical challenge itself (e.g.\ fraud detection or intruder detection).

Multiple statistical approaches exist for anomaly detection, which can
be grouped under the nature of the problem to be solved. \emph{Supervised
  anomaly detection} requires a test dataset with labelled instances
of anomalous entries. Classification machine learning algorithms are
usually employed in this setting, where we consider that we are
provided with an unbalanced dataset~\cite{Lee2018TrainingCC} (we expect a lower fraction of
anomalies). \emph{Unsupervised anomaly
  detection} assumes that no labels exists for the anomalous
data~\cite{eskin2002geometric}. Clustering approaches are usually
employed in this setting, assuming that most of the data is normal,
and that instances not fitting the distribution of the rest of the
data are anomalous. Finally, \emph{Semi-supervised anomaly detection}
approaches assume that training data is normal, builds a model over
it, and tests the likelihood of a new instance being generated from
the learnt model~\cite{gornitz2013toward}.
\justify
{\bf Detection}: Anomaly detection is in principle harder than missing
data detection, since anomalies are not normally encoded explicitly in the
data. Approaches to anomaly detection can be classified as
\emph{univariate} (based only on an individual feature) or
\emph{multivariate} (based on multiple features).  At a univariate
level, a cell might be anomalous for \emph{syntatic} reasons, e.g.\
that the observed value \q{Error} is not consistent with the known
type (e.g.\ integer) of the variable\footnote{ {In the database
  literature this is termed a \emph{domain constraint}.}}
\cite{ceritli2019ptype}.  Another cause of anomaly is
\emph{semantic}, e.g.\ that for an integer field representing months
only the values 1-12 are valid, and hence an entry 13 would be
anomalous. A further class of anomaly is \emph{statistical}, where
a model is built for the clean distribution of a feature, and
anomalies are detected by being low-probability events under the clean
distribution.

For multivariate statistical anomaly detection, one should distinguish between
row outlier detection (where a whole data instance is declared to 
be an outlier, and discarded), and cell outlier detection, where only
one or more cells in the row are identified as anomalous\footnote{Our
  terminology here follows \cite{eduardo2019robust}.}.
Some classical algorithms for unsupervised row outlier detection
include Isolation Forests~\cite{liu2008isolation} and One Class
SVMs~\cite{chen2001one}. More recently deep learning models have been
used for cell outlier
detection, see e.g. \cite{zhou2017anomaly},\cite{eduardo2019robust}.

A multivariate approach to anomaly detection from the databases
  community makes use of  rule-based \emph{integrity constraints}, such as
functional dependencies or conditional functional dependencies,
see e.g.\
  \cite{ilyas2015trends}. One example of such a constraint is that a
  postal code should entail the corresponding city or state, and one
  can check for inconsistencies.  Any violation of an integrity
  constraint in the data would be identified as an error that needs to
  be repaired.


\textit{Practical example}: A univariate statistical approach was followed in~\cite{bjorkman-myers-smith-etal-18} to remove anomalous entries in the Tundra dataset. The observations were grouped by a certain taxon (e.g.\ species, genus or site) and then a threshold was chosen based on eight times the standard deviation of a trait under the taxon ($8\sigma$), flagging as anomalous every observation that deviated from the mean of the taxon more than the threshold. This procedure was employed for all the traits except \q{plant height} and \q{leaf area}, based on the recommendations of the domain experts.
\justify
{\bf Repair}: anomaly repair techniques are similar to missing data repair. Normally, data scientists either remove the anomalous entries completely or, when locating anomalous values under a feature, they substitute those values by other \emph{sensible} values.

\textit{Practical example}: in the Tundra dataset, the anomalous entries detected with the approach mentioned above were removed from the data, in order to avoid introducing biases into the model.


\subsubsection{Non-Stationarity}
\label{sec:non_stat}

This issue involves detecting any \emph{change} occurring in the data
distribution over some period of time. 
This problem is usually referred as change point detection (CPD) in time
series analysis, see e.g.~\cite{basseville1988detecting}. 
However, non-stationarity problems can also arise in tabular data. In particular, a well known problem in machine learning is \emph{dataset shift}, occurring when the distribution of the input data differs between training and test stages~\cite{quionero2009dataset}. Additionally, a common source of non-stationarity in wrangling problems arises from a change in how the data is collected.
In some cases, the protocol of data collection can change over time,
e.g.\ the units of some measurements are changed, labels can be
recoded, etc. In other cases, when a clear data collection protocol is
not established, different people or institutions may collect data using
different criteria, and a final combined dataset presents clear
format variability that remains constant over some chunks of the data
(normally related to the person or institution that collected that
piece of information).
%
\justify
{\bf Change points}: CPD in time series is usually
concerned in identifying those points in time where a specific change
in the probabilistic distribution of the data
occurs~\cite{basseville1988detecting}. Changes may occur
in the mean, variance, correlation or spectral density, among others. Also,
a change can be defined as an abrupt change in the distribution or a
drift over time over the probability distribution. We refer
to~\cite{aminikhanghahi2017survey} for a review on CPD methods and~\cite{van2020evaluation} for an extensive evaluation of CPD algorithms.
\justify
{\bf Protocol changes}: this problem involves any process that changes
the collection of the data over a period of time. This problem is
clearly related with the problem of table unioning described in Section~\ref{sec:dataint}.
In many cases
data scientists are provided directly with a dataset that contains
a combination of different tables. It is common that this combination results in the format of some features changing, new labels being added to the data~\cite{camilleri2019extended} or measurements being recorded
in different units (see Section~\ref{sec: canon}). Detecting when these changes in the data collection occur is vital to solve other wrangling
problems arising from these changes.


\textit{Practical example 1}: In the CleanEHR, a patient undergoing a specific test was always encoded with \q{1}. However, a test not being performed on a patient had two different encodings: \q{0} for some groups of patients and \q{NaN} for other groups. These two sets of encodings follow a sequential pattern, since the data is an aggregate of datasets from different hospitals and each
hospital used different encoding conventions (either \q{1} and \q{0}, or \q{1} and \q{NaN}).
Without noticing this problem, a data scientist might remove some of
these features from the data, assuming that NaN implies missing data.


\textit{Practical example 2}: In the Broadband dataset, the feature indicating whether the broadband connection was in a rural or urban environment not only had different names across tables, but also some tables encoded only \q{Rural} and \q{Urban} (installments from 2013 and 2014) and others \q{Rural}, \q{Urban} and \q{Semi-Urban} (installments from 2015). In principle, these features could be
combined easily, however, another problem is raised from
the modelling perspective. There is no knowledge on whether in the
\q{Rural-Urban} installments there were not any \q{Semi-Urban} instances, or whether
a new encoding of the different locations was introduced at some point
(non-stationarity problem). The data analyst decided to encode \q{Semi-Urban} instances as \q{Rural}, to get a binary classification problem, since the test data did not have any \q{Semi-Urban} instances either.

\subsection{Feature Engineering}
\label{sec:feat_eng}

Once the data is relatively clean and organized in a single table, a
data scientist will start to be concerned about the representation of
the features in the data needed for the analytical task at hand. It is common that the raw features in the data are not suitable for the machine learning algorithm to be employed. For example, data might contain a combination of string and numerical features, while the method they intend to employ only accepts numerical values as inputs. A \emph{feature engineering} process, where some transformations are applied to the features, or new features are created based on existing ones, is arguably one of the most important aspects of a data scientist's work. This can be carried out based on the analyst's
knowledge or beliefs about the task at hand, and the information
required to solve it. Additionally, methods employed to extract meaningful features from the data are usually domain specific. For example, when the data involves sensor readings, images or other low-level information, signal processing and computer vision techniques may be used to determine meaningful features that can be used. Also, more recently deep learning approaches have been used to \emph{learn} appropriate features in an end-to-end fashion  in such domains, given sufficient training data.

We define {\bf feature engineering} as any manipulation of the data
that involves changing the number of features contained in the data or
their properties. Some common feature manipulations include:
aggregating several features with the same information into a unique
feature, creating new features based on already existing features in
the data, deleting unnecessary features or creating one-hot encoding
representations of categorical variables. Notice that some of these manipulations can also be categorized as table manipulations (see Section~\ref{sec:data_transformation}).

\textit{Practical example 1}: in the HES dataset, redundant
information in the demographics table was combined into a newly
created feature \q{PensionerOnly}, and all the features with
information already contained under \q{PensionerOnly} were removed. Also, \q{House.age} was originally encoded as a string with a range of years
(e.g.\ 1950-1966). To encode this as a numerical feature, the difference between the midpoint of the range and 2007 was computed (which was
the latest age of any house in the dataset).


\textit{Practical example 2}: In the CleanEHR dataset, many features that were directly correlated with the time of death of the patients had to be removed
(e.g.\ \q{CCL3D} recorded the number of days the patient was in the
ICU), as these would \q{leak} information about the chosen target variable.
Additionally, every patient that was admitted
into the ICU already dead and every patient where no time of death was
recorded was removed from the dataset, since the evaluation of the predictive model is not possible without ground truth.

\begin{table*}[t]
\setlength{\tabcolsep}{1pt}
\centering
\caption{Summary of data wrangling problems.
\vspace{-0.35cm}
\label{t:summary}}
\resizebox{\textwidth}{!}{
\begin{tabular}{@{}|p{5cm}p{16cm}|@{}}


\hline
\multicolumn{2}{|c|}{\multirow{2}{*}{\Large{Data Organization}}} \\ 
& \\\hline

\textbf{Data Parsing} & 
What is the original data format? {How can it be read properly?} \\ \hline
\textbf{Data Dictionary} & \\
\hspace{0.2cm} Table description & What is the meaning of the table? Is there a header? Are there several tables in the data? \\
\hspace{0.2cm} Feature description & What is the type of each
feature? What are the possible ranges of values? What is the meaning of the feature?\\
\hspace{0.2cm} Value description & What is the meaning of the
values of a feature? Are there unexpected values in a feature? \\ \hline
\textbf{Data Integration} & \\
\hspace{0.2cm} Table joining & How do we join tables containing the same entities? How do we detect common entities across tables?\\
\hspace{0.2cm} Table unioning & How do we aggregate tables with common information together?  Are common features consistent across tables?\\ 
\hspace{0.2cm} Heterogeneous integration & How do we combine e.g.\ time series with relational data or images? How do we query multiple datasets stored in different locations?\\ 
\hspace{0.2cm} Deduplication & How do we detect duplicate entities in the data?\\ \hline
\textbf{Data Transformation} & \\
\hspace{0.2cm} Table transformations & Should we use a \q{tall} format or
a \q{wide} format? Should we remove rows or columns? \\
\hspace{0.2cm} Information extraction & Are there properties of the data contained in free text features? If so, how do we extract this information?\\ \hline
\hline
\multicolumn{2}{|c|}{\multirow{2}{*}{\Large{Data Quality}}} \\ 
& \\\hline
\textbf{Canonicalization} & \\
\hspace{0.2cm} Cell entity resolution & Do some values in a feature
represent the same concept? If so, how do we group them under a common concept?\\
\hspace{0.2cm} Features & Are there different formats of a feature present? What is the standard format? \\
\hspace{0.2cm} Units & Are measurements recorded with the same units?\\ \hline
\textbf{Missing Data} & \\
\hspace{0.2cm} Detection & Which cells are missing? Which values are used to represent missing data?\\
\hspace{0.2cm} Understanding & Why is the data missing? Is it MCAR, MAR or MNAR? \\
\hspace{0.2cm} Repair & Should we impute missing values? Should we delete rows or columns containing missing values? \\ \hline
\textbf{Anomalies} & \\
\hspace{0.2cm} Detection & How do we identify anomalous entries across features? \\
\hspace{0.2cm} Repair & Should we leave these values in the data or repair them? Should we remove rows with anomalies? \\ \hline
\textbf{Non-Stationarity} & \\
\hspace{0.2cm} Change points & Does the behaviour of a time series change abruptly? Is there a meaning behind these changes? \\
\hspace{0.2cm} Protocol changes & Do the properties of the data change while exploring it sequentially? Are there different feature encodings present across consecutive entities?\\ \hline
\hline
\multicolumn{2}{|c|}{\multirow{2}{*}{\Large{Feature Engineering}}} \\ 
& \\\hline
\textbf{Feature Engineering} & Are the features in the data useful for the task at hand? Do we need to concatenate or expand features? Do we need to create new features based on existing ones? \\ \hline
\end{tabular}
}
\end{table*}

\section{Related Work \label{sec:rel-work}}



The literature on data engineering is extensive with many works
addressing general problems on the field, while others only focusing on a
specific data wrangling issue. We discuss here a number of other
review papers on general data engineering.

The authors in \cite{rahm2000data} divide data cleaning problems into schema level
and instance level.  At the schema level the focus is on integrity
constraints (which fall under our anomaly detection heading), while
the instance level contains problems such as missing data,
canonicalization, data integration, data description, etc. Note that
these instance level issues are not systematically categorized in the
paper, but only listed as examples (see e.g., their Table 2). The authors
also distinguish single-source and multi-source problems, although 
the main new issue that they identify for multiple sources is that of 
record linkage.

The CRISP-DM framework of \cite{crisp-dm-00} highlights two
phases relating to data engineering, which they call data
understanding and data preparation. Their data understanding phase
includes data description  (similar to our data 
dictionary heading), and verification of
data quality (which relates to our headings of missing data and 
anomalies), as well as data collection and data exploration. Their
data preparation phase includes data selection and cleaning tasks
(including the handling of missing data), data integration, and
tasks we describe under feature engineering including 
the formatting of data for modelling tools, and constructing 
derived features. Our descriptions in Section\ \ref{sec:class} 
are more detailed than those provided in the CRISP-DM paper.

A taxonomy of \q{dirty data} is provided in \cite{kim2003taxonomy}.
At the highest level this taxonomy is split between missing data (their
class 1), versus non-missing data (class 2). Class 2 is further broken
down into 2.1 ``wrong data'' (due e.g.\ to violation of integrity
constraints, and to failures of deduplication), and 2.2 ``not wrong,
but unusable data'' (which include issues of canonicalization and
record linkage). In our view the choice of a top-level split on
missing/non-missing data creates a rather unbalanced taxonomy.  Note
also that the authors do not provide much discussion of the topics
falling under our \emph{data organization} heading.

In \cite{Castelijns2019abc}, the authors refined the notion of \emph{data readiness
  levels} introduced by \cite{lawrence2017data}, to provide levels C
(conceive), B (believe) and A (analyze). Level C relates mainly to our
data organization issue, especially data parsing and data
integration. Their level B mainly covers topics falling under our data
quality heading (missing data, canonicalization) along with data
dictionary and data integration (deduplication) topics from data
organization. Their level A includes mainly feature engineering
issues, but also includes outlier detection (which we class under
anomaly detection) and a semantic description of features (see our
data dictionary heading). Note also that our paper provides a 
fuller discussion of the issues; in \cite{Castelijns2019abc} the
topics mentioned under levels A, B and C get only one paragraph each.


In \cite{abedjan2015profiling}, the authors survey the task of \emph{data profiling},
which is defined (following \cite{johnson-09}) as ``the activity of
creating small but informative summaries of a database''. This is
clearly related to our data dictionary task, but they also
discuss detecting multivariate dependencies, which we 
mainly cover under the heading of anomaly detection.

%

The identification of various problems as described above leads to the
question of tools to help addressing them. We have pointed out various
tools to address specific problems in Section \ref{sec:class}. However,
there is also a need for systems that allow the \emph{integration} of
multiple individual tools to guide data scientists during the
%
data engineering process. \cite{Serban:2013} provide a survey of
intelligent assistants for data analysis. One prominent (commercial)
system is Trifacta \cite{trifacta18}, which is based on preceding systems
such as Wrangler \cite{kandel2011wrangler}. It provides a
system for interactive data transformations that suggests applicable
transformations based on the current interaction with the
data \cite{heer-hellerstein-kandel-15}.
Wrangler provides common data transformations such as mapping
operations, fold operations, sorting or aggregations among
others. It also infers statistical data types and
higher-level semantic types. Another tool is \textsc{Data Civiilzer}
\cite{data-civilizer-18} which offers functionality including
entity matching across multiple heterogeneous sources, handling missing
values, and canonicalization.

\section{Case studies \label{sec:examples}}


We showcase here the full data wrangling analysis performed in our
four case studies. For each dataset X, we start by giving a description
of the raw dataset, followed by: Section \ref{sec:examples}.X.1 gives
a description of the analytical challenge associated with the dataset. Section \ref{sec:examples}.X.2 provides an ordered pipeline of the data engineering issues addressed in each dataset to obtain a clean version of the data. Section \ref{sec:examples}.X.3 describes the issues that arose against the headings of our classification. Finally, Section \ref{sec:examples}.X.4 describes the model employed for the associated analytical task and the
final results of the analysis.


\subsection{Tundra Traits \label{sec:Tundra}}

This use-case comprises two separate datasets:\\
1) The \emph{Tundra Traits Team database}. It contains nearly 92,000 measurements of 18 plant traits. The most frequently measured traits (more than $1\,000$ observations each) include
   plant height, leaf area, specific leaf area, leaf fresh and dry mass, leaf
   dry matter content, leaf nitrogen content, leaf carbon content, leaf
   phosphorus content, seed mass, and stem specific density. The dataset also comes
   with a cleaning script prepared by its originators, which we have followed.\footnote{Publicly available at https://github.com/ShrubHub/TraitHub. We were kindly granted early access to this dataset by Isla H. Myers-Smith and Anne Bjorkman of the sTUNDRA group (https://teamshrub.wordpress.com/research/tundra-trait-team/)}\\
2) The \q{CRU Time Series} temperature data and
   rainfall
   data\footnote{http://catalogue.ceda.ac.uk/\\uuid/58a8802721c94c66ae45c3baa4d814d0}. These
   are global datasets, gridded to $0.5º$. We use only the date range 1950 to 2016.

\subsubsection{Analytical Challenge}

The challenge posed in~\cite{bjorkman-myers-smith-etal-18} was to infer how climate affects shrub growth across a variety of measured traits. We attempt to fit a model predicting certain traits (e.g.\ vegetative height) from temperature, similar to the first panel of Figure 2b in~\cite{bjorkman-myers-smith-etal-18}. Much of our approach follows~\cite{bjorkman-myers-smith-etal-18}; however, we have made some simplifications and assumptions, and we acknowledge that we are not experts in this domain. We also do not have access to the original code.

There is some terminology to know. The measurement subjects are individual
plants, and the types of measurements on a plant are called
\emph{traits}. For example, \q{vegetative height} is a trait. Only the species of a plant
is recorded, so no identification of the same plant across different measurements of the same trait is possible; however, multiple traits on the same plant are measured and recorded together. The experimental protocol in this field, as we understand it, is to delineate, for each observational \emph{site}, a number of individual \emph{plots} (or \emph{subsites}) and to record all plants within each plot, typically in a single session on a particular day. Thus, an \q{individual observational unit} is a single plant, within a particular plot, on a given day; the properties of the unit are the traits.
Notice that the creators of the Tundra dataset provided a script (\q{TTT\_data\_cleaning\_script.R}) that attempts to make the raw data more consistent and interpretable, although they do not perform a complete data wrangling. \\

\subsubsection{Pipeline of data engineering steps}

The first five steps listed below were carried out using the
early-access version of the data made available privately to
us. The remaining steps apply to the publicly available 
version of the data (what they call data\_raw).\\
1) Loading the Tundra Traits data (DP):
	\begin{itemize}
	\item The csv of the data in wide format is loaded (DP).
	\item An individual ID variable is created for each plant (FE).
	\item Header names are renamed into more informative names (DD).
	\end{itemize}	
2) Data is transformed from a wide format into a tall format, where
  two variables are created: a \q{Trait} variable with the names of
  the traits and a \q{Value} variable with the corresponding values
  for each trait (DT).\\
3) A removal of untrustworthy data is performed at this stage, according to different rules (DT):
	\begin{itemize}
	\item Some observations are removed because the analysts were unsure about the units of the measurements.
	\item Some traits were removed because the analysts were unsure of what they meant.
	\item They removed plants according to their \q{Treatment} feature values.
	\end{itemize}
4) A data dictionary process occur at this point, where additional
  information is added into different features (DD):
	\begin{itemize}
	\item Units are added for each trait (they create new features).
	\item New names for the traits are added (they create additional features).
	\end{itemize}
5) A \q{Genus} feature is created, using only the first part of the \q{AccSpeciesName} feature in the data (FE).\\
6) At this stage, the analysts create different summary statistics from the data according to different groupings, and use them to filter part of the observations in the dataset according to different rules (DT + FE).

The list of groupings considered in the analysis are: By Trait, by Trait and \q{AccSpeciesName}, by Trait and Genus, by Trait, \q{AccSpeciesName} and \q{SiteName}, by Trait, Genus and \q{SiteName}.

		
The list of statistics computed in a given group, and added for each individual $x_i$ in that group are:
		\begin{itemize}
		\item Median, standard deviation and number of elements of the group.
		\item Mean of the rest of the elements of the group, without considering $x_i$.
		\item Error risk of each value under the group, according to the computed a) mean and standard deviation and b) median and standard deviation. Error risk is explained further under Anomaly Detection in sec.\ \ref{sec:tundra-engg}.
		\end{itemize}
7) After this filtering process, all unnecessary features are removed from the dataset (most of the previously computed statistics) obtaining a cleaned version of their data (DT). This is the state where, in the publicly available version of the
data, the authors say that the data is clean. Note that the integration of the temperature data from the CRU dataset is not performed in the public repository.\\
8) Temperature data from the CRU dataset is integrated into the Tundra dataset according to the location of the plants (longitude and latitude) and the day and year when the measurements were recorded (DI).

\subsubsection{Data Engineering Issues \label{sec:tundra-engg}}


{\bf Data Dictionary}: some issues were raised during the exploration of the data.\\
1) Latitude and longitude coordinates are given for each row in the
  Tundra dataset, but it is not stated whether those coordinates apply to a site or to a
plot. Inspection of the data seems to indicate that in fact the coordinates
apply at site level, so that all plots within a site should be taken to have the same coordinates.\\
2) In the original analysis~\cite{bjorkman-myers-smith-etal-18}, the authors define temperature as \q{the average temperature of the warmest quarter of the year}. For simplicity our analysis fixes a particular month and uses the average temperature in that month.\\
3) Columns were renamed to a standard terminology.\footnote{Provided by \q{TTT\_data\_cleaning\_script.R}}\\
4) Units for each measurement were added.\footnote{Also provided by \q{TTT\_data\_cleaning\_script.R}}
\justify
{\bf Data integration}: Our model
requires pairs of observations, consisting of individual plant
traits, $y_i$, and the temperature corresponding to the plot $T_\Delta$. There
are two immediate problems: (i) The CRU temperature data does not come with
plot labels; (ii) We need an average
environmental temperature, not necessarily the monthly average temperature
available in the CRU dataset.
The CRU data is gridded: the region of interest is divided into cells whose
centres are $0.5^\circ$ apart, starting at $-179.75^\circ\text{N},
-89.75^\circ\text{E}$ and ending at $179.75^\circ\text{N}, 89.75^\circ\text{E}$.
Each file (apart from its header rows) gives the temperature (or precipitation)
for a specific month, at a matrix of geographical locations. Thus, to determine
the monthly temperature for a given plot, we must load that month's temperature
data into a matrix, convert the row and column indices into longitude and
latitude, and look up the coordinates of a particular plot.
\justify
{\bf Data Transformation}: The original format of the Tundra dataset had one row for each observation of an individual plant,
   in which each possible measurement was represented by a column. However, since most observations only include a subset of the possible measurements, there were many missing values in this format. So the table has been changed from \q{wide} to \q{tall} format: missing measurements are now simply not represented.\footnote{This was done following \q{TTT\_data\_cleaning\_script.R}}
\justify
{\bf Canonicalization}: there was an underlying cell entity resolution problem in the dataset were the same plants were recorded with different name variations (including typos). It was necessary to group the rows referring to the same plants and to provide a unique identifier for each plant (\q{AccSpeciesName} column).
\justify
{\bf Missing Data}: For some entries in the Tundra dataset, a temperature measurement is not available. There were two sources of missing data (1) the geographical location or the year of observation were not recorded for a specific measurement, and as such, a temperature value could not be extracted during the data integration and (b) a temperature value for a specific geographical location and year of observation was not available in the \q{CRU Time Series} dataset.

In both cases, the value in the matrix is replaced by a code meaning \q{missing}.
The metadata header rows in each file specify the value of disguised missing ($-9999.999$). However, the value that is actually used in the data is not the one given; it is $-10000.0$.
\justify
{\bf Anomaly Detection}: There is significant identification and removal of \q{anomalous} data in this
task. Some of it was done by the sTUNDRA team before releasing the dataset; additional filtering was done for the journal article. We have used the initial filtering exactly as given\footnote{provided by \q{TTT\_data\_cleaning\_script.R}} but our secondary filtering may be slightly different from the authors'.

Both kinds of filtering are described in the \emph{Methods} section of 
\cite{bjorkman-myers-smith-etal-18}
and both follow a similar strategy: (i) group the observations by a certain
taxon (which may be species, genus, site, or particular interactions of these);
(ii) choose a threshold \q{error risk} (multiples of the standard deviation of a
trait within the taxon); (iii) remove observations further from the mean of the
taxon than the threshold "error risk". (For certain taxon, the threshold error
risk depends on the count of observations.) 
For example, observations more than $8\sigma$ from the mean of all observations of a particular trait are excluded (unless the trait is plant height or leaf area). 

Presumably, these tactics reflect domain knowledge about how datasets like these become corrupted (for example, perhaps incorrect assumptions about units are common), although a principled strategy is not clear.
\justify
{\bf Feature Engineering}: All covariates were \q{centred} by subtracting their species mean.\\

\subsubsection{Modelling}

After this preliminary clean-up, the analyst has a dataset of $66308$
observations covering 18 traits and 528 species. For the specific challenge
addressed in~\cite{bjorkman-myers-smith-etal-18}, the authors perform additional filtering,
mostly to remove observations that would otherwise be difficult to integrate into the modelling. 

We note that the line between \q{wrangling} and \q{analysis} is
becoming rather blurred at this stage. In particular, one could imagine a
modelling approach that allowed certain observations to have missing components, whilst using the data in the other components.
We followed~\cite{bjorkman-myers-smith-etal-18} by:

\begin{enumerate}
\item Removing all observations without a geographical location (latitude and longitude), year of observation or temperature data;
   
\item Removing all observations for species present in strictly fewer than four observation sites;
   
\item Removing all sites with fewer than 10 observations;

\item Removing species without a sufficient temperature span in their observations, where \q{sufficient} is defined by \q{those species for which traits had been measured in at least four unique locations spanning a temperature range of at least 10\% of the entire temperature range.}
\end{enumerate}


In this analysis, we used a modified version of the hierarchical model described in~\cite{bjorkman-myers-smith-etal-18}, with some changes in the notation. Let $\Sigma$ range over species, $\Delta$ range over plots, and let
individual observations be identified by $i$. We use the following hierarchical model of the relationship between the measurement of a plant trait, $y$, and temperature, $T$:

$$ 
\begin{aligned}
  &\log y_i(T)\mid \Sigma, \Delta \sim 
    \operatorname{Normal}(\alpha_\Sigma + \beta_\Sigma T_\Delta, \sigma) \\
  &\alpha_\Sigma \sim 
    \operatorname{Normal}(\mu_A, \sigma_A),
  \beta_\Sigma \sim
    \operatorname{Normal}(\mu_B, \sigma_B)
\end{aligned} 
$$
with parameters $\alpha_\Sigma$, $\beta_\Sigma$ and $\sigma$ and hyperparameters  $\mu_A$, $\mu_B$, $\sigma_A$ and $\sigma_B$, to be inferred from the data.

Different Bayesian models can now be fitted to the data, the analyst used
Stan\footnote{https://mc-stan.org/} to perform this step. A variety of goodness of fit and convergence
tests were performed: examination of paired plots, of trace plots, density
sampled estimates vs original data, leave one out checks. The model described
above was found to perform best among alternatives, as established by its $R^2$
(coefficient of determination) calculated over the test set of $0.75$.\\

{\bf Results}: This model can be evaluated using goodness of fit diagnostics and held-out data. We split our dataset in two: train (80\% of the total observations) and test (20\% of the total) and assess the model by predicting trait values on the test set.
For every data point of the test data, which consists of a species and a temperature for a given trait, 1000 parameter values were sampled from the fitted model sampling chains. These values were used to produce 1000 trait estimates as per the model
described above. The mean trait value from these estimates was then taken as the prediction, and compared to ground truth.\\

\begin{figure}
    \centering
    \caption{Trait \q{Leaf nitrogen (N) content per leaf dry mass (mg/g)} true observations compared with predicted observations from the test set. The ideal actual-predicted line is shown in red.}
    \includegraphics[width=1.0\linewidth]{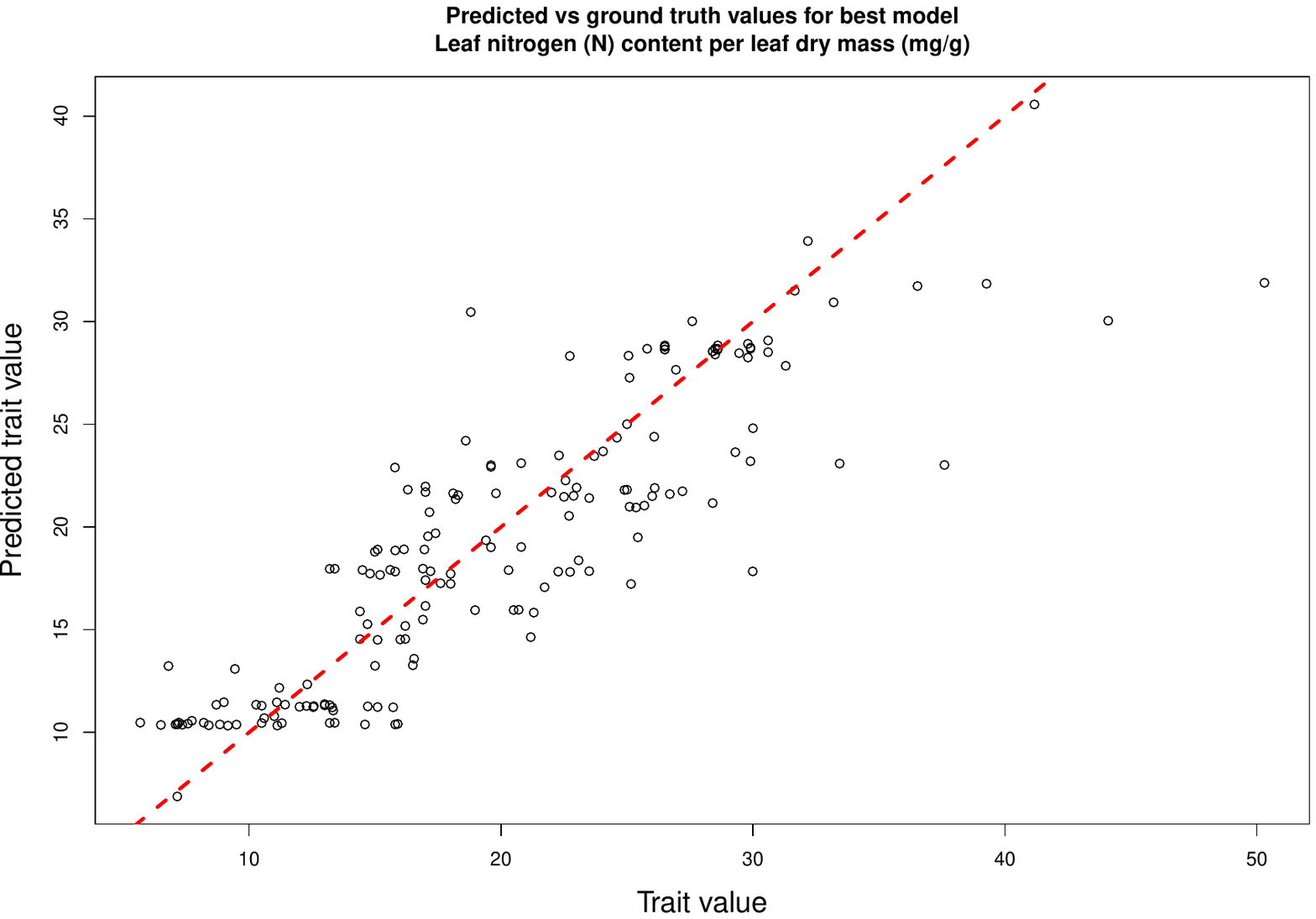}
   
    \label{fig:refeeding}
\end{figure}

Figure~\ref{fig:refeeding} shows how observations sampled from the fitted model for \q{Leaf nitrogen (N) content per leaf dry mass} capture at least part of the effects of temperature. The model is partially accurate in predicting trait values given site temperature and species. We omit the precipitation data, although the wrangling and modelling phases are identical and can be accomplished with no modifications to the code besides a change in the CRU dataset in use.


%


\subsection{Household Electricity Survey \label{sec:HES}}

The Household Electricity Survey 2010--2011 (Department of Energy and Climate Change 2016), a study commissioned by the UK government, collected detailed time-series measurements of the electrical energy consumption of individual appliances across around 200 households in England. The data from that study are available from the UK Data Archive (UKDA, but only on application); an overview is publicly available from the government's website~\cite{HES}.

The available data from the archive are not the raw measurements but have instead been \q{cleaned} by Cambridge Architectural Research Ltd (CAR). Fortunately, CAR have also provided a report describing the process of data cleaning that they undertook. In the following, we are reporting the relevant information from the provided documentation.

All households (all owner-occupied) were monitored from 2010 to 2011. Periodic measurements were taken of the power consumption of all electrical appliances in each household. The majority of households were monitored for a single month. The power drawn by appliances in these households was recorded every two-minutes. The remaining households were monitored for a full year, with measurements made every 10 minutes (every two minutes in some cases). Finally, the study also collected qualitative \q{diaries} from each household, in which the occupants recorded how they used each appliance. We have not made any use of these diaries here.   

Within the data are three primary entities:
\begin{itemize}
\item \emph{Households}: one of the 250 residences for which monitoring took place.  
\item \emph{Appliances}: an electrical appliance that was monitored.  
\item \emph{Profiles}: electricity load profile, where the power usage from a given appliance or group of appliances is measured at regular intervals.
\end{itemize}

The dataset consists of 36 CSV files, spread across 11 directories. For our purposes we will group 6 of these directories into one, called \texttt{appgroupdata}. The directories are then as follows:

\begin{itemize}
\item \texttt{small}: Metadata about appliances.
\item \texttt{originalhes}: Unmodified version of the HES dataset.
\item \texttt{experianhes}: Files created by Experian that give a consumer classification for each household.
\item \texttt{appdata}: One file containing metadata relating to the monitored appliances.
\item \texttt{anonhes}: Anonymized HES files containing demographic information and metadata about the houses.
\item \texttt{appgroupdata}: Files containing the profile information for each appliance that was monitored.
\end{itemize}

There are several encodings (or keys) that appear across tables, and are necessary for combining information from multiple tables (joining the tables). These are:

\begin{itemize}
\item \textbf{household id}: a number that corresponds to a specific household for which monitoring took place.  
\item \textbf{appliance code}: a number that corresponds to a specific type of appliance (e.g., \texttt{182 -> trouser press}).  
\item \textbf{appliance type code}: a number that corresponds to a group of appliances (e.g., \texttt{0 -> Cold Appliances}).  
\item \textbf{interval id}: a number that refers to the type of monitoring that occurred for a given appliance (e.g., \texttt{1 -> for one month at two minute intervals}).  
\end{itemize}

There are some other encodings, but the four described above appear most frequently across the tables. The \q{important files} (defined as so by
Cambridge Architectural Research in their report) are as follows:
\texttt{appliance\_codes.csv}, \texttt{appliance\_type\_codes.csv}, \texttt{appliance\_types.csv},\texttt{total\_profiles.csv}, \\\texttt{appliance\_data.csv}, \texttt{rdsap\_public\_anon.csv}, \texttt{ipsos\-anonymized\-corrected\_310713.csv}, 
and the folder \texttt{appgroupdata} containing 6 CSV files with the profile information of each appliance.\\

\subsubsection{Analytical Challenge}

The dataset can be used to investigate disaggregation models, where the signal with the overall electricity consumption is used to estimate the consumption from component appliances. \cite{zhong2014interleaved} provides an approach to the full disaggregation problem using a hidden Markov model. Since the focus of the present analysis is data wrangling, this task is overly involved for our purposes. Instead, the analyst attempted to solve a simpler problem. The different appliances in the HES data are categorised into \q{Appliance Groups}. Examples of appliance groups are \q{heating}, \q{audiovisual} and \q{cold appliances}. There are 13 such groups in total. The analyst attempted to predict the consumption of a given household in each of these appliance groups over a whole day (instead of at 2 or 10 minute intervals), based on the household information, the day and month of the observation and the total electricity consumption for that day.

\subsubsection{Pipeline of data engineering steps}

\begin{enumerate}[leftmargin=\parindent]
\item Loading the necessary recording files information:
	\begin{itemize}
	\item Parsing the csv files (DP)
	\item Selecting the necessary files among all folders (DD)
	\item Adding header names to the profile data (DD)
	\item Removing white spaces from header names (CA)
	\end{itemize}

\item Record linkage where we add the appliance type code of each appliance into the profile information (DI). This appliance type code is referred as \q{GroupCode} in the different files.

	\begin{itemize}
	\item Three files are linked: \q{appliance\_type\_codes.csv}, \q{appliance\_types.csv} and the profile data.
	\item \q{appliance\_type\_codes.csv} contains \q{GroupCode} and \q{Name}, \q{appliance\_types.csv} contains \q{ApplianceCode} and \q{GroupCode}, and the profile data contains the \q{ApplianceCode} for each measurement.
	\item First, each \q{ApplianceCode} is linked to a \q{Name} using the \q{GroupCode} as primary key.
	\item Two names for \q{GroupCode} are \q{Other} and \q{Unknown}, so every \q{ApplianceCode} related to one of these appliances groups is removed (MD).
	\item Lastly, each measurement in the profile data is linked to a \q{GroupCode}, using \q{ApplianceCode}  as the primary key. Notice that, by removing the \q{ApplianceCode} linked to \q{Other} and \q{Unknown}, in the final integrated data, those \q{GroupCode} missing values are now stardardized to \q{NaN}.
	\end{itemize}
	
\item After the DI step, we remove all rows that contain missing
  values in the \q{GroupCode} feature (DT + MD). These values correspond to temperature readings instead of power consumption readings.


\item Loading the demographic data (DP) and selecting only a subset of ten features from the data (DD + DT).

\item Now, different data wrangling challenges are tackled in the demographic data, to obtain a cleaner version:
	\begin{itemize}
	\item From \q{HouseholdOcuppancy}, transform any \q{6+} code into \q{6} (CA).
	\item Create \q{PensionerOnly} indicator feature using \q{SinglePensioner} and \q{MultiplePensioner} (FE).
	\item Remove redundant features (DT).
	\item \q{Social.Grade} is an ordinal feature (e.g. A, B, C1, etc) where each category is transformed into an ordered numerical value and the only missing entry is replaced with the mode of the feature (FE + MD).
	\item \q{House.age} is transformed from a year interval (e.g. 1930-1949) into a numerical value, using the difference between the year where the measurements were taken (2010) and the middle value of the interval (FE).
		\begin{itemize}
		\item \q{-1} indicates unknown, so these values are replaced with the mode of the feature (MD).
		\item \q{2007 onwards} was encoded as \q{3} (2010 - 2007).
		\end{itemize}
	\end{itemize}
	
\item Creation of data $X$ and target $y$ variables:
	\begin{itemize}
	\item The target $y$ is the aggregate power consumption per \q{groupCode} for each household and date (FE+DT).
	\item The data $X$ is created through a two step process:
		\begin{itemize}
		\item Aggregate the total consumption for each household and date (DT).
		\item Link the demographic data into this aggregate data using the Household ID as primary key (DI).
		\end{itemize}
	\end{itemize}
	
\item Feature engineering processes done on both $X$ and $y$:
	\begin{itemize}
	\item In $X$, the month and day of the week are
          extracted from the date, creating 2 new features: \q{month}, a number between 1 and 12 and \q{day of the week}, a one-hot encoded variable with 7 categories (FE).
	\item Three Fourier components are created based on the month of the recording for each household and date (FE). Afterwards, \q{month}  is removed (DT).
	\item For the target, a standard $\log(1+x)$ transformation for non-negative target variables is applied for each \q{GroupCode} power consuption feature (FE).
	\item Numerical features are scaled (both in $X$ and $y$), dividing them by the difference between their min and max values (FE).
	\end{itemize}

\end{enumerate}

\subsubsection{Data Engineering Issues}

%
It is worth noting that many more wrangling challenges are described in the user guide for this dataset. It appears that the wrangling described in the user guide has been carried out already, so that the version available for download from UKDA has been pre-processed.
\justify
{\bf Data Dictionary}: The tables in this dataset form a rich relational structure that must be inferred by inspection of the files and using the documentation. It is sometimes unclear what the meaning of a given field is, or even the purpose of a sometimes redundant table. Example of the former: the \texttt{Index} column in the \texttt{monitered\_appl.csv} file. Example of the latter: \texttt{appliance\_attributes.csv} is actually redundant but this is not obvious immediately. 
Presently, this wrangling challenge is solved by spending a substantial amount of time inspecting the data, possibly missing relevant information in the process.
\justify
{\bf Data Integration}: In may of the tables, fields that are in fact referring to the same information are given different names. This makes the task of joining tables difficult. For example, the household identifier is sometimes given field name \texttt{Household} (e.g., in \texttt{appliance\_data.csv}), but on other occasions is referred to as \texttt{Clt Ref} (e.g., in \texttt{rdsap\_glazing\_details\_anon.csv}). This is an example of a key identification problem during during record linkage.
This challenge is again solved manually by the analyst, with the obvious costs and risks of introducing errors. An automated agent inspecting whether two columns might actually contain values from the same domain would greatly ease this inspection (primary key - foreign key matcher).
\justify
{\bf Data Transformation}: Subsets of observations in the dataset were manually identified and removed due to different problems. For example, some of the rows in the monitoring data did not
correspond to electricity readings, but instead outdoor temperature
measurements.
\justify
{\bf Canonicalization}:  Many examples of this problem exist in the data. For example, the \texttt{Refrigerator\_volume} column in \texttt{appliance\_data.csv} does not have an explicit unit specified in the field name. Different entries specify different units (e.g.\ litres, cubic feet or no unit specified), which can themselves be written in different ways (e.g.\ \q{Litres}, \q{L}).

Another canonicalization problem is related with some columns expected to be numeric that are not so. For example \texttt{Win Area} in \texttt{rdsap\_glazing\_details\_anon.csv} is not numeric because windows with area greater than 10 (the unit is not specified) have values \q{10+}. To use this column, therefore, one would have to modify the entries that are \q{10+}.
The same problem appeared in the \texttt{HouseholdOccupancy} column, which was encoded as a string, and households with six or more tenants were encoded as \q{6+}. To use this as a numerical feature, the column was cast to integer type and set \q{6+} to \q{6}.

The analyst here acted on a case-by-case basis, often introducing ad-hoc rules to uniform variable names, unit of measures, as well as transforming the data as needed.
\justify
{\bf Missing Data}: Missing entries appear in multiple tables.\\ Example:
\texttt{ipsos-anonymized-corrected\_310713.csv} has some missing entries in the survey
question fields. The various diary files also have a reasonable fraction of
missing values.
%
Also, the encodings of the missing values vary across files and, more importantly, within the same attribute. For example, in the table \texttt{diary\_dish\_washer.csv}, the \texttt{Options} column can take values \texttt{NULL}, \texttt{N/A} or  \texttt{N/a}, all of which mean that data is not available.

The analyst here acted on a case-by-case basis.
For example, the  \texttt{Social.grade} column contained a single
missing value, which was imputed with the most common value in the
remaining rows. In the monitoring data, there was a significant number
of day/house combinations for which the aggregated consumption for a
given appliance group was zero. It remains unclear whether this should
be considered as a genuine reading, or treated instead as missing or
bad data. This was the biggest wrangling challenge for the present use
case. 
\justify
{\bf Feature Engineering}:
the following original features were used during the analysis:

\begin{itemize}
\item \texttt{TotalUsage}: the total electricity consumption of a household on a given day.
\item \texttt{HouseholdOccupancy}: the number of people living in the house.
\item \texttt{HouseholdWithChildren}: a binary indicator, set to `1` if any children live in the house.
\item \texttt{Social.grade}: the "social grade" of the individuals living in the house.
\item Day of the week of the observation, one-hot-encoded.
\item Month of the year.
\end{itemize}

Two features were added after performing some feature engineering to put them into a proper format for our model: 

\begin{enumerate}
\item There was redundant information in the demographics table. The analyst created the \texttt{PensionerOnly} feature, and removed columns that contained information already present in other columns.
\item \texttt{House.age} was encoded as a string, e.g.\ \q{1950-1966}. To encode this as a numerical feature, the difference between the midpoint of the range and 2007 was taken. There was a single missing value in this column, which was set to \q{-1}. This was manually identified and imputed with the most common age range in the column.
\end{enumerate}

\subsubsection{Modelling}

The analyst then turned to the analytical challenge as follows. A
baseline linear model performed poorly. Upon further inspection,
this was found to be because a significant fraction of the electricity
readings in each appliance group were zero: a linear model struggled
to model these data. It is unclear whether the readings of zero are problems with data or correct measurements. Additionally, for some appliance groups, the frequency of zero
readings was sufficiently high that the analyst simply dropped the
appliance group, resulting in only 7 categories being predicted.

Afterwards, for each appliance group, the analyst modelled the consumption as follows:
\begin{enumerate}
\item Classify the house/day combination as \q{zero} or \q{not zero}, using a random forest classifier.
\item For the combinations classified as \q{not zero}, use a linear regression to predict the actual consumption.
\end{enumerate}
This approach worked more effectively than the original approach (linear regression only). Additionally, we used the following approach for each appliance group as a baseline:
\begin{enumerate}
\item For each house/day combination in the holdout set, draw a binary indicator from a binomial distribution with parameter $p$ set to the fraction of non-zero readings for this appliance group in the training set.
\item For house/day combinations where the binary indicator is 0, set the predicted consumption to zero.
\item For house/day combinations where the binary indicator is 1, set the predicted consumption to the mean consumption of non-zero readings in the training set.
\end{enumerate}
The comparison of the baseline to the model is shown in Table~\ref{t:hes-summary}. The model produces a smaller mean absolute error than the baseline for each appliance group, but is not a particularly good fit to the data.
\begin{table}[h]
\setlength{\tabcolsep}{2pt}
\centering
\caption{Comparison of baseline with model predicted energy consumption of
appliance types. Overall, there were $1\,546$ data points.}
\vspace{0.3cm}
\label{t:hes-summary}
\begin{tabular}{lcc}
\hline
Appliance Group             & Baseline MAE & Model MAE \\
\hline
Showers                     & 0.150 & 0.187\\
Heating                     & 0.124 & 0.116\\
Water heating               & 0.098 & 0.111\\
Washing/drying/dishwasher   & 0.374& 0.237\\
Audiovisual                 & 0.407 & 0.228\\
Cooking                     & 0.103 & 0.078\\
Cold appliances             & 0.070& 0.061\\
\textbf{Overall}                 & \textbf{0.189} & \textbf{0.145}\\
\end{tabular}
\end{table}



\subsection{Critical Care Health Informatics Collaborative}
\label{sec:cleanEHR}

The CleanEHR anonymized and public dataset can be requested online. It contains records for 1978 patients (one record each) who died at a hospital (or in some cases arrived dead), with 263 fields, including 154 longitudinal fields (time-series). These fields cover patient demographics, physiology, laboratory, and medication information for each patient/record, recorded in intensive care units across several NHS trusts in England. The full dataset, available only to selected researchers, contains 22,628 admissions from 2014 to 2016, with about 119 million data points. The dataset comes as an R data object, which can be most profitably accessed with an accompanying R package \footnote{Information on how to get the dataset and an introduction to it can be found in here: https://cran.r-project.org/web/packages/cleanEHR/vignettes/cchic\_overview.html. The R package is available at https://github.com/ropensci/cleanEHR. There is also a blog post on how to use it: https://ropensci.github.io/cleanEHR/data\_clean.html. A detailed explanation of each field is found in https://github.com/ropensci/cleanEHR/wiki under Data-set-1.0 and CCHIC-Data-Fields}.

We consider the dataset called `anon\_public\_d` containing 1978 patient episodes. It is anonymized and only contains fatal outcomes. Some examples of the non time-series fields (demographic) found in the dataset are ADNO (Critical care local identifier), HCM (Height) and RAICU1	(Primary reason for admission to your unit).
The longitudinal fields are time-series variables, that correspond to physiological/nursing measurements, laboratory tests results and drugs administered to the patient during their stay. Some examples are mean arterial blood pressure (Physiology), sodium (Laboratory) and adrenaline (Drugs).


\subsubsection{Analytical Challenge}

The analytical challenge we first set ourselves is to predict the amount of time it takes for a patient to die
(in minutes), after they arrived at the hospital alive. Unfortunately, the proposed model fitted the data poorly, so we decided to shift the analysis into predicting which patients die in the first 100 hours from admission to the unit.

\subsubsection{Pipeline of data engineering steps}

\begin{enumerate}
\item First the analyst loads both CSVs for the demographic data and the time series statistics data (DD) and creates an extended table with the features of both tables using \q{ADNO} as primary key (DT).
\item Data transformations and feature engineering steps are done to the data with the analytical task in mind.
	\begin{itemize}
	\item Remove all rows where the date of death, time of death or date of access to the ICU were not recorded (DT + MD).
	\item Remove features fully missing in the data (MD).
	\item Standardize dates and times into the same format (CA), creating a numerical variable based on those dates (FE).
	\item A \q{survival class} variable is created, where patients are divided into two groups according to whether they survived for more or less than 100 hours.
	\item All patients who died before arriving at the ICU were removed from the analysis (DT).
	\end{itemize}

\item Creation of a numerical age feature based on other features in the data (FE).

\item Categorical variables are transformed into one-hot variables (FE). In particular, \q{RAICU1}, \q{RAICU2} and \q{OPCHM} features are simplified, and only the first two components are used (e.g. 2.1.2 $\rightarrow$ 2.1)

\item Further rows and features were removed from the data according to different criteria (DT).
	\begin{itemize}
	\item All patients that didn't live for for than 10 hours were removed from the study (DT). 
	\item Previous versions of transformed features were removed.
	\item Other variables were removed because they were heavily correlated to the target variable. For example, \q{CCL3D} recorded the number of days the patient was in the ICU. Other variables recorded the number of days the patients were under any type of support system. These variables disclose information about the number of days that the patients were alive, so they were removed from the analysis.
	\end{itemize}
	
\item Missing data (MD): a) Features with more than 50\% of their values missing were deleted. b) For the remaining features, missing values were replaced by \q{-1}.

\end{enumerate}

\subsubsection{Data Engineering Issues}

There was a previous integration process done before getting access to the data we are employing in our analysis.
The data coming from
eleven adult ICUs at five UK teaching hospitals is passed through a
pipeline that extracts, links, normalizes and anonymizes the EHR
data. The resulting data is then processed using the cleanEHR toolkit, covering the most common processing and cleaning operations for this type of data (see above). 
Some of the most important functions in the cleanEHR package converts the various asynchronous lists of time-dependent measurements into a table of measurements with a customizable binning. A second function is used to join together separate but sequential critical care admissions into a unified illness spell~\cite{harris2018critical}.
\justify
{\bf Data Dictionary}: The analyst started by attempting to understand the available data fields, whose contents are highly domain-specific and for which documentation is limited. Column names like DOB, WKG, HDIS or DOD were not readily understood by the data scientist. A matching process of each feature name in the dataset and a website containing the
descriptions and information of the different features was necessary to understand the data.\footnote{https://github.com/ropensci/cleanEHR/wiki under Data-set-1.0 and CCHIC-Data-Fields}
\justify
{\bf Data Integration}: the analyst loaded the two datasets (the demographics and the time-series analysis) and linked them into a single extended table using the ADNO (admission number) field.
%
%
Since the primary key (ADNO) was present in both tables, the mapping was done directly by adding the features in the time-series table to the demographics table.
\justify
{\bf Data Transformation}: after the DI process, it was necessary to remove rows in the data according to two criteria: a) all rows without date and time of admission or death were removed, as their absence makes the predictive task impossible and b) all patients admitted dead (i.e. whose elapsed time to death is zero or negative) were removed.
\justify
{\bf Canonicalization}: Several fields are not standardized. For example, datetimes show a variety of encodings being used, such as: `1970-01-01T00:01:00`, `1970-01-01` and `00:01:00` (i.e. datetime, date and time). The analyst had to deal with these issues case-by-case.
\justify
{\bf Missing Data}: Several fields contain missing values. This is to be expected since many measurements are only performed in case of medical necessity. Unfortunately, an added challenge is that the analyst is never sure if a value is missing due to the measurement not being performed or due to other reasons. We expect the former case to be the most common, yet we do not know in practice.

Intuitively, NaN values usually entail a lack of observation
due to a measurement deemed not important for the
patient/condition at hand. Thus, it was decided to drop
all columns containing more than 50\% of NaN values (including some completely empty columns). The remaining NaN
values were imputed with -1 to have a clear indication that this particular value was not necessary or unobserved (the default value of 0 was first considered, however there are fields where zero can be the result of a measurement). A more detailed analysis of missing values and of different imputation strategies might improve results.
\justify
{\bf Non-Stationarity}: while exploring the data the data scientists noticed that consecutive patients could be grouped under different groups according to the properties of some of the attributes. For example, when looking at HCM (height of the patients), the patients with IDs between 800-1300 had their height measured in metres, while the rest of the patients had their height measured in centimetres. 
According to the documentation, this attribute should have been recorded in centimetres, so a proper transformation from metres to centimetres for this attribute was necessary (related to canonicalization). We are aware that most of this non-stationarity problems are related to the preliminary integration script provided by the dataset.
 \justify
{\bf Feature Engineering}:
The analyst performed the following set of operations in sequence:\\
1) Export date-times from all datetime columns, using a variety of string representations (`1970-01-01T00:01:00`, `1970-01-01` and `00:01:00`. This format variability operation was performed by defining a set of rules that transformed all datetime strings into datetime Python objects (which internally abide to ISO standards).\\
2) Calculate the elapsed time to death (prediction target or dependent variable) by subtracting the time of admission from the time of death.\\
3) Calculate the age from birth and admission dates.\\
4) Transform the taxonomies used for diagnoses into dummy one-hot encoded variables (`RAICU1`, `RAICU2` and `OCPMH` above).\\
5) Create dummy one-hot encoded variables for all remaining categorical variables.\\
6) Extract relevant statistics from time-series data. These were the mean, standard deviation, and last value measured (but more variables could be used in the future).\\
7) Drop unnecessary columns (either judged as not predictive or duplicated with dummy variables).\\
8) Drop columns that can leak the time of death  (e.g.\ number of days on the ICU `CCL3D`), this is a challenging step since it is not always clear when the data was recoded.\\

\subsubsection{Modelling}

Here it is important not to include variables that can leak asynchronous information from the target (time to death). For example, using the full time-series data from the patient 
to build the descriptive variables would be misleading, because in a real predictive scenario this corresponds to future information that will not be available.
Taking this into consideration, only the first 10 hours of time-series data was used to build the model, and predictions are done only in patients that have lived at least those 10 hours. These choices were made in order to include as much of the data as possible, without making the model overly complex.
%

Two datasets were obtained from the previous wrangling steps. A first sample containing only the demographic variables and a second one with both demographic and time-series information. The sample containing both
demographic and time-series data only has patients that lived at least 10 hours. The demographic only sample has all patients that arrived alive to the unit.
After the data wrangling process, the demographic-only sample had 176 variables for 1730 patients. The demographic plus time-series data had 373 variables corresponding to 1586 patients. Both samples are divided in training and testing samples with a ratio of 80/20 and two models are built based on the datasets.
A linear regression model with ElasticNet regularization (joint L1 and L2)~\cite{zou2005regularization} was implemented to predict the time (in minutes) elapsed from admission to death. However, no suitable model was found. The regression model fits the data poorly, reporting an R-squared value of 3\% for the demographics only sample and 7\% for the full dataset.

The analysis focus was shifted to predict which patients die in the first 100 hours from admission to the unit. Random Forests~\cite{breiman2001random} were used for classification. 
Both samples are split into two classes: patients that died in the first 100 hours and patients that survived for 100 hours. For the demographic-only sample 51\% of the patients correspond to the first class (and 49\% to the second).
The demographic plus time-series sample is divided in 47\% and 53\% for the first and second class, respectively.    
 
The sample with demographic-only data achieves an average classification accuracy of 53\%, which is barely better than the
classification baseline of predicting the most frequent class (51\% based on the class balance for this sample).
This result hints that this sample might not contain enough information to accurately classify 
patients that die in the first 100 hours from admission (see Table \ref{t:final_class_eval_demographic}). 
The classification accuracy obtained by using the demographic plus time-series data (where all patients survived at least 10 hours and their previous time-series data
is used in the model) is shown in Table \ref{t:final_class_eval_demographic_plus_timeseries}. The average accuracy for this sample value is of
66\%, which is a significant improvement compared to the demographic-only sample and the classification baseline of predicting the most frequent class (53\%). This result shows that the 
time-varying fields bring important information to the system, which can be of great use in the training of a more
sophisticated survival model in the future. 
One of the main challenges of this analysis is to know how and when
some of the demographic variables are measured or computed. For
example, the apache score or AMUAI variables are good predictive
features but they can be asynchronous to the information available up
to 10 hours after admission. In this model those features are not used, but they might be useful to increase the performance or build a more sophisticated model. 

\begin{table}[h]
\setlength{\tabcolsep}{2pt}
\centering
\caption{Confusion matrix of classification of time of death within the first 100 hours from admission (demographic only sample).}
\label{t:final_class_eval_demographic}
\vspace{0.3cm}
\begin{tabular}{lcc}
\hline
Time of death  (\%)     & $<$ 100 hours (Pred.)  & $>$ 100 hours (Pred.) \\
\hline
$<$ 100 hours (True) & 0.38                 & 0.62                \\
$>$ 100 hours (True) & 0.30                 & 0.70                 \\
\hline
\end{tabular}
\end{table}
\begin{table}[h]
\setlength{\tabcolsep}{2pt}
\centering
\caption{Confusion matrix of classification of time of death within the first 100 hours from admission (demographic plus timeseries sample).}
\label{t:final_class_eval_demographic_plus_timeseries}
\vspace{0.3cm}
\begin{tabular}{lcc}
\hline
Time of death  (\%)     & $<$ 100 hours (Pred.)  & $>$ 100 hours (Pred.) \\
\hline
$<$ 100 hours (True) & 0.75                   & 0.25                 \\
$>$ 100 hours (True) & 0.33                  & 0.67                 \\
\hline
\end{tabular}
\end{table}

\subsection{Ofcom Consumer Broadband Performance}
\label{sec:broadband}

The UK’s Office of Communications (Ofcom) conducts an annual survey of the
performance of residential fixed-line broadband services\footnote{https://data.gov.uk/dataset/uk-fixed-line-broadband-performance}. These data are
published as a table in annual instalments: a new spreadsheet each year
containing that year's data. Each row of this table corresponds to a recruited broadband user panellist. There
are $1\,450$ rows in the 2013 dataset, $1\,971$ rows in the 2014 data, and
$2\,802$ rows in the 2015 data.

\subsubsection{Analytical Challenge}

There isn't a natural challenge accompanying this data. We chose the following simplistic challenge:
fit a simple model to detect whether a broadband speed corresponds to an urban or rural area based on several features from the data. We will use the annual installments from 2014 and 2015 to train our model and the installment from 2016 to test it.

\subsubsection{Pipeline of data engineering steps}

1) There are two installments from 2014 (May and November), one from 2015 and another from 2016 that needs to be loaded (DP). Some of these tables have notes at the beginning of the csv, so we need to carefully select the part of the csv that corresponds to the data.\\
2) Every table has different features and different names for the same features. We  rename those features into a standard naming and identify the correct features (DD). Since only 6 features were deemed necessary to the analysis, we only rename those and delete the rest from the tables (DT).\\
3) Installments from 2014 and 2015 are concatenated into the same table (DI).
\begin{itemize}
\item The feature indicating whether a broadband device is in an urban or rural area changes between 2014 and 2015. A protocol change occurred in 2015, where they introduced the term \q{Semi-Urban} to indicate an area that is in between urban and rural (NS). The analyst decided to encode \q{Semi-Urban} instances as \q{Rural}, since there were fewer instances of \q{Rural} in the data.
\item There are several duplicates across 2014 and 2015, since some devices were measured several times at different months. The analyst decided to keep the records from the most recent measurements, removing the rest of the duplicates (DI, deduplication).
\end{itemize}	
4) The analyst removed from the test set (2016) any broadband devices included both in the training and test set, using the \q{Id} of the device (DI).\\
5) Since the amount of missing data across features was less than $1\%$, the analyst decided to remove entries with missing values, to avoid biasing the model later (MD).\\
6) Training and test datasets were prepared (FE):
	\begin{itemize}
	\item The \q{Urban/Rural} feature was recoded, assigning $0$ to rural and semi-urban areas and $1$ to urban areas (FE).
	\item The \q{Id} feature was deemed unnecessary and removed from the data (DT).
	\item Real features were standardized (FE).
	\end{itemize}


\subsubsection{Data Engineering Issues}


{\bf Data Parsing}: Some of the CSV files contained notes in the first lines of the headers, and automatic parsing failed. The analyst needed to explore the files and identify the positions of the tables in the files to load the data properly.
\justify
{\bf Data Dictionary}: The columns in the data describe characteristics of the geographical region
(such as whether it is rural or urban); the nature of the broadband service (for
example, the provider); and measured characteristics of the broadband service
(for example, the average download speed). There are roughly five fields
describing properties of the region; three fields describing the service; and
around 20 fields describing measured characteristics. As is often the case,
this dataset is not in "normal form" but instead can usefully be thought of as a
relational join of multiple tables. It is left to the analyst to determine the
schema.
\justify
{\bf Data Integration}: Since the data comes in annual installments, it was necessary to aggregate all the tables into an extended table. Unfortunately, a common schema for the features employed during the data collection was not present, so the analysts needed to match corresponding features across tables together and do the integration manually.
Some  challenges they needed to address were:
\begin{enumerate}
\item Reaming columns to a standard name
\item Ordering columns across tables.
\item Not all features appeared in all installments.
\item Recoding some categorical features, so as to use the same encoding in all installments (CA).
\end{enumerate}
Additionally, it was necessary to remove duplicate devices measured across different installments.
\justify
{\bf Missing Data}: There were some features with missing data across installments. The analyst needed to check every feature and treat the missing data accordingly. In this case, since the missing values were less than $1\%$ per feature, instances with missing values were removed from the data.


\subsubsection{Modelling}

For this task, we are trying to detect whether a broadband measurement
comes from an urban or rural area based on the download and upload speed of the broadband, the latency and the web pages loading speed. These features were recorded as an average of the speed during a period of 24 hours. We will make a comparison between different classifiers
for this specific task, training and validating the models with the data from 2014
and 2015, and testing the results in the 2016 data.

We compare the performance of Logistic Regression (LR), Naive Bayes
(NB), SVM and Random Forests (RF). A grid search
for the best set of hyperparameters for each model was performed, using cross-validation with 3 folds on the training set.
%
Table~\ref{t:broadband_classifiers} shows the precision, recall, F1 score and accuracy on the 2016 dataset for the classifiers trained on the 2014 and 2015 data. All the classifiers obtain similar results, with RF performing slightly better.
\begin{table}[h]
\setlength{\tabcolsep}{2pt}
\centering
\caption{Precision, recall, F1 score and accuracy for the different classifiers}
\label{t:broadband_classifiers}
\vspace{0.3cm}
\begin{tabular}{ccccc}
\hline
Model & Precision & Recall & F1 score & Accuracy\\
\hline
NB                  &  0.8031 & 0.8975 & 0.8477  & 0.7666              \\
LR & 0.7933 & 0.9347 &  0.8582  & 0.7764             \\
SVM & 0.7718 & 0.9578 &0.8547 & 0.7644\\
RF & 0.7822 & 0.9774 & 0.8686 & 0.7859\\
\hline
\end{tabular}
\end{table}

\section{Conclusions \label{sec:conc}}
In this paper we have identified three high-level groups of data wrangling
problems, those related with obtaining a proper representation of the
data (data organization), those related to assessing {and 
improving} the quality of
the data (data quality), and feature engineering issues, which heavily
depend on the task at hand and the model employed to solve it.
Furthermore, we have presented the full analysis of four use cases,
where we have provided a systematic pipeline for each of the datasets
to clean them while identifying and classifying the main problems the
data scientists faced during the wrangling steps. We hope that
this work helps to further explore and understand the field of data
engineering, and to value a part of every data scientist's work that most
of the time goes unnoticed both in research and
industry. Additionally, we would like to encourage practitioners to
provide their raw data and the scripts necessary to clean it in order
to advance the field.
In future work we would like to study data engineering workflows across
multiple datasets in order to identify (if possible) common 
structures concerning the ordering of the various wrangling operations. We note
that there can be feedback loops in the process, as described e.g.\ in
\cite{crisp-dm-00}.


%

%

%
 \section*{Acknowledgments}

This work was supported in by The Alan Turing Institute under the EPSRC grant EP/N510129/1 and by funding provided by the UK Government's Defence \& Security Programme in support of the Alan Turing Institute. 
 
We would like to thank the AIDA team including Taha Ceritli, Jiaoyan Chen, James Geddes, Zoubin Ghahramani, Ian Horrocks, Ernesto Jimenez-Ruiz, Charles Sutton, Tomas Petrick and Gerrit van den Burg for helpful discussions and comments on the paper.




%
%
%

\bibliographystyle{IEEEtran}
\bibliography{autods_ckiw}

\end{document}